\documentclass[aps,pre,twocolumn,,showpacs,floatfix]{revtex4}
\usepackage{color}
\usepackage{graphicx}
\usepackage{amssymb,amsfonts,amsmath}

\newcommand{\refeq}[1]{~(\ref{#1})}

\newcommand{\ignore}[1]{}

\begin{document}

\title{Stability of Boolean and continuous dynamics} 
\author{Fakhteh Ghanbarnejad and Konstantin Klemm}
\email{{fakhteh,klemm}@bioinf.uni-leipzig.de}
\affiliation{Bioinformatics Group, Institute for Computer Science,
University of Leipzig, H\"artelstra\ss{}e 16-18, D-04107 Leipzig, Germany
}

\begin{abstract} Regulatory dynamics in biology is often described by
continuous rate equations for continuously varying chemical concentrations. 
Binary discretization of state space and time leads to Boolean dynamics. 
In the latter, the dynamics has been called unstable if flip perturbations
lead to damage spreading. Here we find that this stability classification
strongly differs from the stability properties of the original
continuous dynamics under small perturbations of the state vector.
In particular, random networks of nodes with large sensitivity yield stable
dynamics under small perturbations.
\end{abstract}



\maketitle


The functioning of organisms on the molecular level is a research topic of
increasing attention. Survival and reproduction requires an autonomous
regulation of chemical concentrations in the living cell.  Modeling such
regulatory dynamics, various mathematical approaches have been studied, from
discrete to continuous methods, from deterministic to stochastic techniques,
from static to dynamical models, from detailed to coarse grained perspectives
\cite{Bornholdt:2005}, see ref.~\cite{deJong:2002} for an overview.

Boolean dynamics
\cite{Kauffman:1969,Charteret:2008,Albert:2008,drossel-review,Helikar:2008,Sahoo:2010} is a
framework for modeling regulatory systems, especially for precise sequence
control as observed in morphogenesis \cite{Albert:2003} and cell cycle dynamics
\cite{Li:2004} but also in the regulation of the metabolism \cite{Samal:2008}.
Using binary (on/off) concentrations as an idealization, Boolean dynamics
directly implements the logical skeleton of regulation. Values of system
parameters such as binding constants, production and degradation rates etc.\
are not needed. This abstraction simplifies computation and analytical
treatment. Boolean networks have been extracted directly from the literature
\cite{Helikar:2008,Davidich:2008} of known biochemical interactions or obtained
by discretization of differential equation models \cite{Davidich:2008b}. Known
state sequences and responses of several systems have been faithfully
reproduced by the discrete models \cite{Albert:2003,Li:2004}.

Despite these benefits, modelers do not employ Boolean dynamics as widely as
ordinary or delay differential equations. The latter are embedded in an
established framework for state-{\em continuous} dynamical systems
\cite{Strogatz:1994} which itself builds on the mathematical foundations of
linear algebra and infinitesimal calculus. In particular, the definition of {\em
stability} of a solution  under {\em small} perturbations is based on the
consideration of infinitesimally small neighborhoods in state space. Stability
checks for solutions of the dynamical equations are a salient part of
mathematical modeling. Unstable solutions are not expected to be observed in a
real-world system. 

In the state-{\em discrete} Boolean dynamics, {\em large} perturbations are
normally implemented as a {\em flip}, where the state of a single Boolean
variable is inverted. Then the evolution of the damage is tracked. The damage is
the difference between the state of the perturbed and the unperturbed system.
The return map of the expected size of the damage is known as Derrida plot
\cite{Derrida:1986}. Numerous studies have elucidated the effect of flip
perturbations on regulatory dynamics with Boolean states 
\cite{Kauffman:2003,Shmulevich:2003,Kauffman:2004,Rohlf:2007,
Pomerance:2009,Peixoto:2010}. When asking if a gene-regulatory system reproduces
a prescribed trajectory despite noise, large perturbations are to be considered
in the case of low copy numbers of regulatory molecules and bursty stochastic
response \cite{Eldar:2010}. Small perturbations, however, are more appropriate
when modeling systems with large copy numbers and an integrative response to
filter out bursts, see e.g.\ \cite{Lestas:2010}.

Here we find that the clear distinction between the two types of perturbations
is crucial. In a continuous system, stability or instability under small
perturbations is not indicative of the effect of flip perturbations. Likewise,
probing a Boolean system with flip perturbations does not necessarily provide
information about the stability of the continuous counterpart under small
perturbations.


An $n$-dimensional Boolean map $f:\{0,1\}^n \rightarrow \{0,1\}^n$ gives
rise to a time-discrete dynamics
\begin{equation} \label{eq:boolmap}
x(t+1) = f(x(t)) 
\end{equation}
with $x=(x_1,\dots,x_n)$ being a Boolean state vector (bit string) of $n$
entries. Such a map is equivalent to a {\em Boolean network}. When $f$ is
pictured as a network, a node corresponds to a coordinate $i$ of the Boolean
state vector and a directed edge $j \rightarrow i$ (from node $j$ to node $i$)
is present if the Boolean function $f_i$ explicitly depends on the $j$-th
coordinate.

Let us now define a continuous dynamics whose discretization readily
leads to the Boolean map in Eq.\refeq{eq:boolmap}. Taking values
$y_i(t) \in [0,1]$, $i \in \{1,\dots,n\}$, $t \in \mathbb{R}$, the states
evolve according to the delay differential equation
\begin{equation} \label{eq:ddeours}
\dot{y_i}(t+1) = \alpha \mathop{\mathrm{sgn}}(\tilde f(y(t)) - y_i(t+1))
\end{equation}
with $\alpha$ an inverse time constant. For large $\alpha$, this is essentially Boolean dynamics with fast but
continuous switching between the saturation values.
The simplest choice is  $\tilde f  = f \circ \Theta$ with $\Theta$ the
component-wise step function, $\Theta_i(y) = 1$ if $y_i \ge 1/2$ and
$\Theta_i(y)=0$ otherwise. This choice of continuous dynamics is in close
correspondence with the discrete dynamics in the following sense. Suppose $x(0),
x(1), x(2), \dots$ is a solution of Eq.\refeq{eq:boolmap}. Let
$y(t)$ be a solution of Eq.\refeq{eq:ddeours} such that there is a time
interval $[t_1,t_2]$ with $y(s) = x(0)$ for all $s \in [t_1,t_2]$. 
Then for all future times $t \in \mathbb{N}$ and all $s \in [t_1,t_2]$
\begin{equation}
x(t) = y(\beta t + s) 
\end{equation}
with $\beta = 1+ 1/(2\alpha)$. 
The closest resemblance between Boolean and continuous dynamics is obtained 
when choosing the same initial condition, that is $y(s)=x(0)$ for all $s \in
[-1,0]$. Similar correspondence between Boolean maps and ordinary
differential equations has been studied earlier neglecting
transmission delay \cite{Glass:1972} or implementing more complicated
differential equations \cite{Braunewell:2009,Norrell:2007,Norrell:2009,
Gehrmann:2010} compared to Equation\refeq{eq:ddeours}.


{\em Perturbations. ---} Given a map $f$, the evolution of states is uniquely determined by
Eq.\refeq{eq:ddeours} by an initial condition $y(t)$ on a time interval
of unit length, here taken as $[-1,0]=:I$. We restrict ourselves to
initial conditions that do not vary on $I$, $y(t) = y(0)$ for all $t \in I$.
An initial condition with a {\em small} perturbation is generated as
\begin{equation} \label{eq:pert}
y^\prime_i (t) := \epsilon_i (1-y_i(t)) + (1-\epsilon_i) y_i(t)
\end{equation}
for $t \in I$. The perturbation amplitudes are arbitrary numbers
$\epsilon_i\in\; ]0,1/2[$. 
An initial condition with a {\em flip} perturbation is generated as
\begin{equation}
y^!_i (t) := \left\{ \begin{array}{rl}
1-y_i(t) & \text{if }i=l \\
  y_i(t) & \text{otherwise}
\end{array}\right.
\end{equation}
for $t \in I$ and an arbitrary node $l \in \{1,\dots,n\}$.
Note that the total amplitude $\sum_i \epsilon_i$ of a small perturbation may
exceed the unit amplitude of a flip perturbation. A small perturbation produces
small deviations from the original state potentially at each node. 
A flip perturbation concentrates a maximal deviation at a single node.

We say that the system {\em heals} from the perturbation if the dynamics from
perturbed and unperturbed initial condition eventually become the same except
for an arbitrary time lag. Formally, healing from a small perturbation means
that there are $t_0>0$ and $\tau > -t_0$ such that 
\begin{equation} \label{eq:healing}
y(t) = y^\prime(t+\tau)
\end{equation}
for all $t \ge t_0$. Healing from a flip perturbation means that
Eq.\refeq{eq:healing} holds analogously for $y^!$ instead of $y^\prime$. We
define the heal time $t_\text{heal}$ as the smallest time $t_0$ for which this
holds. 


\begin{figure}
\centerline{\includegraphics[width=0.5\textwidth]{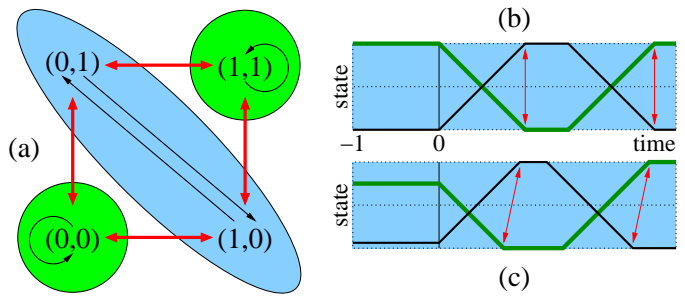}}
\caption{\label{fig:st_graph} (color online). 
Dynamics of two mutually activating nodes.
(a) State space of the Boolean system described by Eq.\refeq{eq:twonodes}.
Thin arrows indicate the mapping $f$ of states by the dynamics, thick bidirectional arrows stand
for flip perturbations. Indicated by shaded areas, the system has three 
dynamical modes (attractors): two fixed points $(0,0)$ and $(1,1)$ and a cycle
of length 2 involving the states $(0,1)$ and $(1,0)$. 
(b) Time evolution of the corresponding continuous system in
Equation\refeq{eq:ddeours} with initial
condition $x_1(0)=1$ (thick curve) and $x_2(0)=0$ (thin curve). 
The two nodes switch in a synchronous mode as indicated by vertical
double arrows akin to the Boolean state sequence $(0,1), (1,0), (0,1), \dots$.
(c) Time evolution from perturbed initial condition, $x_1(0)<1$, $x_2(0)>0$.
The perturbation translates into a phase lag in switching that does
not heal out.
}
\end{figure}

{\em Fixed points and bistable circuits. ---}
Let us first consider a fixed point as the simplest dynamical
behaviour. A fixed point of the continuous dynamics is a state
vector $y^\ast$ such that constant $y(t) = y^\ast$ is a solution
of Eq.\refeq{eq:ddeours}. This in turn means that the time derivative
vanishes at all times, equivalent to $y^\ast = f(y^\ast)$. The fixed points
of the continuous dynamics are exactly the fixed points of the discrete
map $f$. A small perturbation to a fixed point $y^\ast$
always heals, because values after
applying the treshold $\Theta$ remain unchanged, $\tilde f(y^\prime(t))=y^\ast$ 
for all $t \in I$. All fixed points are stable under small perturbations.
However, a flip perturbation to a fixed point does not always heal.
The {\em bistable switch} is an example. Consider a two-dimensional map $f$
with $f(x_1,x_2) = (x_2,x_1)$. It gives rise to the dynamics
\begin{equation} \label{eq:twonodes}
x_1(t+1) = x_2(t) \qquad x_2(t+1) = x_1(t)
\end{equation}
with fixed points $(0,0)$ and $(1,1)$. After perturbing a fixed point by
flipping one node's state, the system does not return to the fixed point. It
remains in the set of the state vectors $(0,1)$ and $(1,0)$ constituting a limit
cycle, cf.\ Figure~\ref{fig:st_graph}(a). The stability of the fixed points is
not obtained when probing the dynamics with flip perturbations.  The bistable
switch constitutes a first simple example of systems with different
stability properties under flip and small perturbations.

In the continuous counterpart of the alternating Boolean state $(0,1)$ and
$(1,0)$, small perturbations do not heal, see Figure~\ref{fig:st_graph}(b,c).  
The effect of a small perturbation is to induce a phase lag in the
oscillation, being discussed in earlier work
\cite{Klemm:2005a,Klemm:2005b,Braunewell:2007,Braunewell:2009}.


\begin{figure}
\centerline{\includegraphics[width=0.48\textwidth]{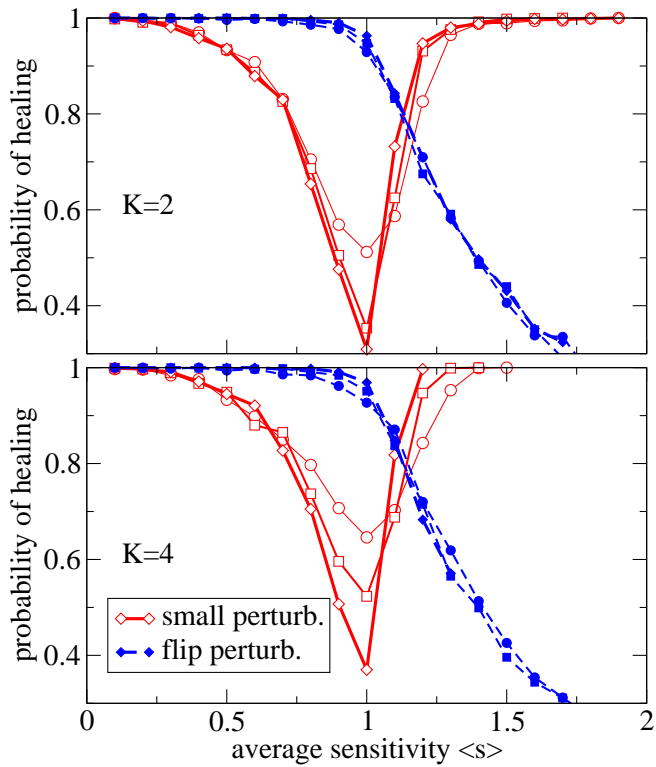}}
\caption{\label{fig:pcon_0} (color online).
Stability of dynamics in random networks under perturbation by spin flip
(dashed) and under continuous perturbation (solid lines) in random networks with
$K=2$  and $K=4$ inputs per node.  Symbols distinguish system size $n=300$
($\circ$),  $1000$ ($\Box$) and $3000$ ($\diamond$). Each data point gives the
relative frequency of healed out perturbations on a set of $10^4$ independent
random realizations of network, initial condition and perturbation. Each
amplitude $\epsilon_i$ of a small perturbation is drawn independently from the
uniform distribution on an interval $[0;r]$ with $0<r<0.5$.  The results are
independent of the choice of $r$. As a general invariance of the dynamics of
Equation~\eqref{eq:ddeours} with $\tilde f  = f \circ \Theta$, the qualitative
effect (healing or spreading) of a small perturbation is not altered when the
amplitude vector is multiplied with a positive scalar keeping each
amplitude $\epsilon_i<0.5$.
}
\end{figure} 

{\em Stability in random networks.---} We now compare the effects of the two
types of perturbations on dynamics in randomly generated networks. An ensemble
of random Boolean networks (RBN) \cite{drossel-review} is defined by the
number of nodes $n$, the number of inputs $K$ of each node, and the probability
distribution of Boolean functions $\pi(f)$. The latter is taken as a maximum
entropy ensemble $\pi_\lambda(f) \propto \exp(\lambda s(f))$ under a given
average sensitivity $\langle s \rangle$. The sensitivity $s(f)$ of a Boolean
function $f$ is the number of flips at one of the $K$ inputs that lead to a
change of the output value, averaged over all input vectors
\cite{Shmulevich:2004}. The resulting value $s(f)$ lies in the range from zero
(for a constant function $f$) to $K$, obtained for a parity function where 
for all input vectors, a flip of a single input state flips the output.
For RBN, where the $K$ inputs of each node are drawn randomly and independently
from the set of $n$ nodes, the average sensitivity $\langle s \rangle$ is the
crucial parameter determining the system's response to flip perturbations
\cite{Shmulevich:2004}. In the limit $n\rightarrow \infty$, these perturbations
heal in ensembles with $\langle s \rangle <1$; they spread when $\langle s
\rangle >1$. This change of behaviour in dependence of $\langle s \rangle$
is reproduced in Figure~\ref{fig:pcon_0} (dashed lines) for varying $K$ and
$n$.

As our main result, we show in Fig.~\ref{fig:pcon_0} that the $\langle s
\rangle$-dependence of the
healing probability under flip perturbations is qualitatively different from
that under small perturbations. Only in the so-called critical region of
$\langle s \rangle\approx 1$, small perturbations spread. Both for $\langle s
\rangle\ll 1$ and $\langle s \rangle\gg 1$, the healing probability tends
towards 1.
This effect is enhanced by increasing system size. In the limit of $n
\rightarrow \infty$ one may expect a finite probability of non-healing only at
$\langle s \rangle= 1$. Then the dynamics is almost always stable under small
perturbations.

\begin{figure}
\includegraphics[width=0.48\textwidth]{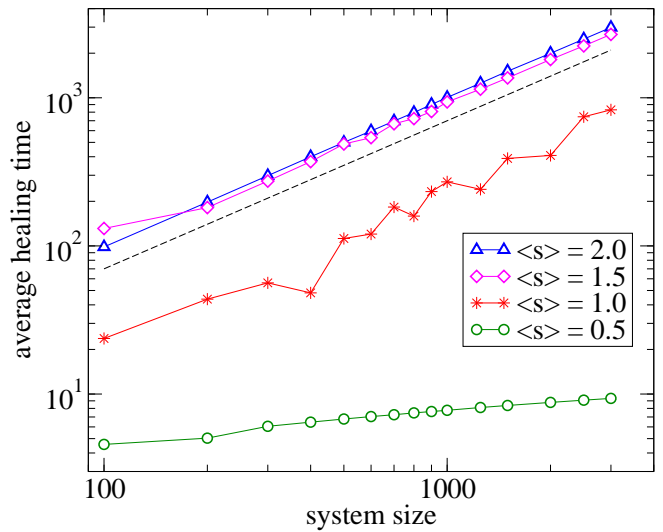}
\caption{\label{fig:healtime} (color online).
The average time to heal from a small perturbation increases linearly with the
number of nodes in the system for sensitivity $\langle s \rangle \ge 1$, and
sublinearly otherwise. The dashed line has slope 1 in this double-logarithmic
plot. Each data point is the average over $t_\text{heal}$ for the subset of
healing realizations. Realizations of network, initial condition and
perturbation are the same as in Figure~\ref{fig:pcon_0}.}
\end{figure}

The average time $t_\text{heal}$ to heal from small perturbations increases
moderately with system size as shown in Figure~\ref{fig:healtime}. For average
sensitivity above $1$, we observe a linear increase $\langle t_\text{heal}
\rangle \propto n$. For lower values of the average sensitivity, the increase
is sublinear.

\begin{figure}
\centerline{\includegraphics[clip,width=0.48\textwidth]{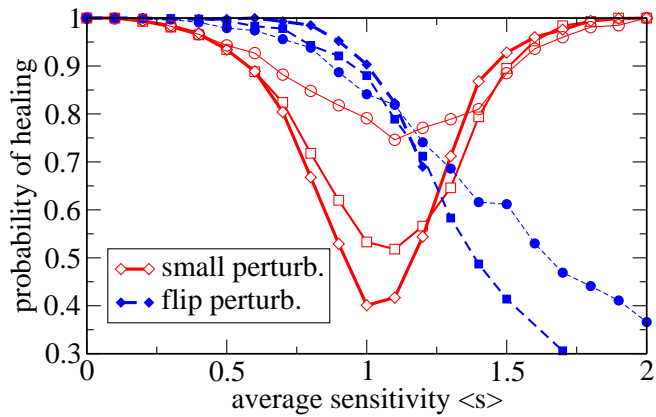}}
\caption{\label{fig:cont_B} (color online).
Healing probabilities remain qualitatively the same
(cf.\ Figure~\ref{fig:pcon_0}) when using the alternative transfer function
$\tilde f_i (y) = \Theta (h_i (y))$ with $h_i(y) = a y_j y_k + b_1 y_j + b_2 y_k + c$;
for node $i$ taking inputs from nodes $j$ and $k$. The parameters
$a,b_1,b_2,c$ are chosen such that $h_i(y) = f_i(y)$ for $y_j,y_k \in \{0,1\}$.
If, for instance, $f_i$ is an AND then $a=1$ and $b_1=b_2=c=0$ so
$\tilde f_i (y) =1$ if and only if the product of inputs $y_j y_k \ge 1/2$.
Each data point is the healing fraction of 1000
realizations of given average sensitivity and system size
$n=30$ ($\circ$), $100$ ($\Box$) and $300$ ($\diamond$).
The perturbation amplitude $\epsilon_i$ is drawn from the
uniform distribution on $[0;0.01]$ independently for each node $i$.}
\end{figure}

The dynamics we have studied so far is simple but not the only possibility to
pass from the Boolean map to a continuous flow. In order to check to what
extent our results depend on this choice we repeat simulations for $K=2$ with
an alternative function $\tilde f$ (cf.\ Equation\refeq{eq:ddeours}) now taking
into account cooperative effects between inputs. Figure~\ref{fig:cont_B} shows
that the same qualitative result obtains under this choice, see figure caption
for details.

In summary, we have shown that the dynamics of large random networks of
switch-like elements typically recovers from small perturbations of the state
vector. Healing is observed naturally at low sensitivity. However, also large
sensitivities of the nodes' functions render the long-term behaviour  of the
whole system insensitive to small perturbations. Instability is observed only in
an intermediate sensitivity regime that shrinks as systems become larger.

The behaviour under small perturbations is essentially different from the
established stability diagram for RBN. Under {\em flip} perturbations, RBN
display a transition from healing to non-healing (damage spreading) behaviour at
average sensitivity $1$. It has been suggested that networks of regulatory
switches position themselves at this transition \cite{Kauffman:1993}, known as
the edge of chaos \cite{Langton:1990}. Then some but not all flip perturbations
spread and therefore allow for complex information processing without rendering
the system unreliable under noise. 

According to our findings, a complementary scenario is worth discussing. The
apparent conflict between responsiveness to external input signals and
resilience to intrinsic noise dissolves when these influences act as
perturbations at separate scales: noise corresponds to small perturbations
whilst input signals are interpreted as the flipping of a state. Under these
assumptions, noise resilience and responsiveness are compatible rather than
conflicting in the regime of average sensitivity above 1. Systems that combine
both beneficial properties are obtained ``for free'' in random networks of
sufficiently sensitive switching elements. 

{\em Acknowledgments.---} The authors thank Gunnar Boldhaus, Florian Greil and
Thimo Rohlf for valuable comments. This work has been financially supported by
Volks\-wagen\-Stiftung through the initiative on Complex
Networks as Phenomena across Disciplines.

\bibliography{paper}

\begin{thebibliography}{34}
\expandafter\ifx\csname natexlab\endcsname\relax\def\natexlab#1{#1}\fi
\expandafter\ifx\csname bibnamefont\endcsname\relax
  \def\bibnamefont#1{#1}\fi
\expandafter\ifx\csname bibfnamefont\endcsname\relax
  \def\bibfnamefont#1{#1}\fi
\expandafter\ifx\csname citenamefont\endcsname\relax
  \def\citenamefont#1{#1}\fi
\expandafter\ifx\csname url\endcsname\relax
  \def\url#1{\texttt{#1}}\fi
\expandafter\ifx\csname urlprefix\endcsname\relax\def\urlprefix{URL }\fi
\providecommand{\bibinfo}[2]{#2}
\providecommand{\eprint}[2][]{\url{#2}}

\bibitem[{\citenamefont{Bornholdt}(2005)}]{Bornholdt:2005}
\bibinfo{author}{\bibfnamefont{S.}~\bibnamefont{Bornholdt}},
  \bibinfo{journal}{Science} \textbf{\bibinfo{volume}{310}},
  \bibinfo{pages}{449} (\bibinfo{year}{2005}).

\bibitem[{\citenamefont{de~Jeong}(2002)}]{deJong:2002}
\bibinfo{author}{\bibfnamefont{H.}~\bibnamefont{de~Jeong}}, \bibinfo{journal}{J
  Comput Biol} \textbf{\bibinfo{volume}{9}}, \bibinfo{pages}{67}
  (\bibinfo{year}{2002}).

\bibitem[{\citenamefont{Kauffman}(1969)}]{Kauffman:1969}
\bibinfo{author}{\bibfnamefont{S.~A.} \bibnamefont{Kauffman}},
  \bibinfo{journal}{J Theor Biol} \textbf{\bibinfo{volume}{22}},
  \bibinfo{pages}{437} (\bibinfo{year}{1969}).

\bibitem[{\citenamefont{Albert et~al.}(2008)\citenamefont{Albert, Thakar, Li,
  Zhang, and Albert}}]{Albert:2008}
\bibinfo{author}{\bibfnamefont{I.}~\bibnamefont{Albert}},
  \bibinfo{author}{\bibfnamefont{J.}~\bibnamefont{Thakar}},
  \bibinfo{author}{\bibfnamefont{S.}~\bibnamefont{Li}},
  \bibinfo{author}{\bibfnamefont{R.}~\bibnamefont{Zhang}}, \bibnamefont{and}
  \bibinfo{author}{\bibfnamefont{R.}~\bibnamefont{Albert}},
  \bibinfo{journal}{Source Code Biol Med} \textbf{\bibinfo{volume}{3}},
  \bibinfo{pages}{16} (\bibinfo{year}{2008}).

\bibitem[{\citenamefont{Drossel}(2007)}]{drossel-review}
\bibinfo{author}{\bibfnamefont{B.}~\bibnamefont{Drossel}},
  \bibinfo{journal}{Reviews of Nonlinear Dynamics and Complexity}
  \textbf{\bibinfo{volume}{1}}, \bibinfo{pages}{69} (\bibinfo{year}{2007}).

\bibitem[{\citenamefont{Helikar et~al.}(2008)\citenamefont{Helikar, Konvalina,
  Heidel, and Rogers}}]{Helikar:2008}
\bibinfo{author}{\bibfnamefont{T.}~\bibnamefont{Helikar}},
  \bibinfo{author}{\bibfnamefont{J.}~\bibnamefont{Konvalina}},
  \bibinfo{author}{\bibfnamefont{J.}~\bibnamefont{Heidel}}, \bibnamefont{and}
  \bibinfo{author}{\bibfnamefont{J.~A.} \bibnamefont{Rogers}},
  \bibinfo{journal}{Proc Natl Acad Sci USA} \textbf{\bibinfo{volume}{105}},
  \bibinfo{pages}{1913} (\bibinfo{year}{2008}).

\bibitem[{\citenamefont{Sahoo et~al.}(2010)\citenamefont{Sahoo, Seita,
  Bhattacharya, Inlay, Weissman, Plevritis, and Dill}}]{Sahoo:2010}
\bibinfo{author}{\bibfnamefont{D.}~\bibnamefont{Sahoo}},
  \bibinfo{author}{\bibfnamefont{J.}~\bibnamefont{Seita}},
  \bibinfo{author}{\bibfnamefont{D.}~\bibnamefont{Bhattacharya}},
  \bibinfo{author}{\bibfnamefont{M.~A.} \bibnamefont{Inlay}},
  \bibinfo{author}{\bibfnamefont{I.~L.} \bibnamefont{Weissman}},
  \bibinfo{author}{\bibfnamefont{S.~K.} \bibnamefont{Plevritis}},
  \bibnamefont{and} \bibinfo{author}{\bibfnamefont{D.~L.} \bibnamefont{Dill}},
  \bibinfo{journal}{Proc Natl Acad Sci USA} \textbf{\bibinfo{volume}{107}},
  \bibinfo{pages}{5732} (\bibinfo{year}{2010}).

\bibitem[{\citenamefont{Carteret et~al.}(2008)\citenamefont{Carteret, Rose, and
  Kauffman}}]{Charteret:2008}
\bibinfo{author}{\bibfnamefont{H.~A.} \bibnamefont{Carteret}},
  \bibinfo{author}{\bibfnamefont{K.~J.} \bibnamefont{Rose}}, \bibnamefont{and}
  \bibinfo{author}{\bibfnamefont{S.~A.} \bibnamefont{Kauffman}},
  \bibinfo{journal}{Phys. Rev. Lett.} \textbf{\bibinfo{volume}{101}},
  \bibinfo{pages}{218702} (\bibinfo{year}{2008}).

\bibitem[{\citenamefont{Albert and Othmer}(2003)}]{Albert:2003}
\bibinfo{author}{\bibfnamefont{R.}~\bibnamefont{Albert}} \bibnamefont{and}
  \bibinfo{author}{\bibfnamefont{H.}~\bibnamefont{Othmer}}, \bibinfo{journal}{J
  Theor Biol} \textbf{\bibinfo{volume}{223}}, \bibinfo{pages}{1}
  (\bibinfo{year}{2003}).

\bibitem[{\citenamefont{Li et~al.}(2004)\citenamefont{Li, Long, Lu, Ouyang, and
  Tang}}]{Li:2004}
\bibinfo{author}{\bibfnamefont{F.}~\bibnamefont{Li}},
  \bibinfo{author}{\bibfnamefont{T.}~\bibnamefont{Long}},
  \bibinfo{author}{\bibfnamefont{Y.}~\bibnamefont{Lu}},
  \bibinfo{author}{\bibfnamefont{Q.}~\bibnamefont{Ouyang}}, \bibnamefont{and}
  \bibinfo{author}{\bibfnamefont{C.}~\bibnamefont{Tang}},
  \bibinfo{journal}{Proc Natl Acad Sci USA} \textbf{\bibinfo{volume}{101}},
  \bibinfo{pages}{4781} (\bibinfo{year}{2004}).

\bibitem[{\citenamefont{Samal and Jain}(2008)}]{Samal:2008}
\bibinfo{author}{\bibfnamefont{A.}~\bibnamefont{Samal}} \bibnamefont{and}
  \bibinfo{author}{\bibfnamefont{S.}~\bibnamefont{Jain}}, \bibinfo{journal}{BMC
  Syst Biol} \textbf{\bibinfo{volume}{2}}, \bibinfo{pages}{21}
  (\bibinfo{year}{2008}).

\bibitem[{\citenamefont{Davidich and
  Bornholdt}(2008{\natexlab{a}})}]{Davidich:2008}
\bibinfo{author}{\bibfnamefont{M.~I.} \bibnamefont{Davidich}} \bibnamefont{and}
  \bibinfo{author}{\bibfnamefont{S.}~\bibnamefont{Bornholdt}},
  \bibinfo{journal}{PLoS ONE} \textbf{\bibinfo{volume}{3}},
  \bibinfo{pages}{e1672} (\bibinfo{year}{2008}{\natexlab{a}}).

\bibitem[{\citenamefont{Davidich and
  Bornholdt}(2008{\natexlab{b}})}]{Davidich:2008b}
\bibinfo{author}{\bibfnamefont{M.}~\bibnamefont{Davidich}} \bibnamefont{and}
  \bibinfo{author}{\bibfnamefont{S.}~\bibnamefont{Bornholdt}},
  \bibinfo{journal}{J Theor Biol} \textbf{\bibinfo{volume}{255}},
  \bibinfo{pages}{269 } (\bibinfo{year}{2008}{\natexlab{b}}).

\bibitem[{\citenamefont{Strogatz}(1994)}]{Strogatz:1994}
\bibinfo{author}{\bibfnamefont{S.~H.} \bibnamefont{Strogatz}},
  \emph{\bibinfo{title}{Nonlinear Dynamics and Chaos: With Applications to
  Physics, Biology, Chemistry and Engineering}} (\bibinfo{publisher}{Westview
  Press, Boulder}, \bibinfo{year}{1994}).

\bibitem[{\citenamefont{Derrida and Pomeau}(1986)}]{Derrida:1986}
\bibinfo{author}{\bibfnamefont{B.}~\bibnamefont{Derrida}} \bibnamefont{and}
  \bibinfo{author}{\bibfnamefont{Y.}~\bibnamefont{Pomeau}},
  \bibinfo{journal}{Europhys Lett} \textbf{\bibinfo{volume}{1}},
  \bibinfo{pages}{45} (\bibinfo{year}{1986}).

\bibitem[{\citenamefont{Kauffman et~al.}(2003)\citenamefont{Kauffman, Peterson,
  Samuelsson, and Troein}}]{Kauffman:2003}
\bibinfo{author}{\bibfnamefont{S.}~\bibnamefont{Kauffman}},
  \bibinfo{author}{\bibfnamefont{C.}~\bibnamefont{Peterson}},
  \bibinfo{author}{\bibfnamefont{B.}~\bibnamefont{Samuelsson}},
  \bibnamefont{and} \bibinfo{author}{\bibfnamefont{C.}~\bibnamefont{Troein}},
  \bibinfo{journal}{Proc Natl Acad Sci USA} \textbf{\bibinfo{volume}{100}},
  \bibinfo{pages}{14796} (\bibinfo{year}{2003}).

\bibitem[{\citenamefont{Shmulevich et~al.}(2003)\citenamefont{Shmulevich,
  L{\"a}hdesm{\"a}ki, Dougherty, Astola, and Zhang}}]{Shmulevich:2003}
\bibinfo{author}{\bibfnamefont{I.}~\bibnamefont{Shmulevich}},
  \bibinfo{author}{\bibfnamefont{H.}~\bibnamefont{L{\"a}hdesm{\"a}ki}},
  \bibinfo{author}{\bibfnamefont{E.~R.} \bibnamefont{Dougherty}},
  \bibinfo{author}{\bibfnamefont{J.}~\bibnamefont{Astola}}, \bibnamefont{and}
  \bibinfo{author}{\bibfnamefont{W.}~\bibnamefont{Zhang}},
  \bibinfo{journal}{Proc Natl Acad Sci USA} \textbf{\bibinfo{volume}{100}},
  \bibinfo{pages}{10734} (\bibinfo{year}{2003}).

\bibitem[{\citenamefont{Kauffman et~al.}(2004)\citenamefont{Kauffman, Peterson,
  Samuelsson, and Troein}}]{Kauffman:2004}
\bibinfo{author}{\bibfnamefont{S.}~\bibnamefont{Kauffman}},
  \bibinfo{author}{\bibfnamefont{C.}~\bibnamefont{Peterson}},
  \bibinfo{author}{\bibfnamefont{B.}~\bibnamefont{Samuelsson}},
  \bibnamefont{and} \bibinfo{author}{\bibfnamefont{C.}~\bibnamefont{Troein}},
  \bibinfo{journal}{Proc Natl Acad Sci USA} \textbf{\bibinfo{volume}{101}},
  \bibinfo{pages}{17102} (\bibinfo{year}{2004}).

\bibitem[{\citenamefont{Rohlf et~al.}(2007)\citenamefont{Rohlf, Gulbahce, and
  Teuscher}}]{Rohlf:2007}
\bibinfo{author}{\bibfnamefont{T.}~\bibnamefont{Rohlf}},
  \bibinfo{author}{\bibfnamefont{N.}~\bibnamefont{Gulbahce}}, \bibnamefont{and}
  \bibinfo{author}{\bibfnamefont{C.}~\bibnamefont{Teuscher}},
  \bibinfo{journal}{Phys Rev Lett} \textbf{\bibinfo{volume}{99}},
  \bibinfo{pages}{248701} (\bibinfo{year}{2007}).

\bibitem[{\citenamefont{Pomerance et~al.}(2009)\citenamefont{Pomerance, Ott,
  Girvan, and Losert}}]{Pomerance:2009}
\bibinfo{author}{\bibfnamefont{A.}~\bibnamefont{Pomerance}},
  \bibinfo{author}{\bibfnamefont{E.}~\bibnamefont{Ott}},
  \bibinfo{author}{\bibfnamefont{M.}~\bibnamefont{Girvan}}, \bibnamefont{and}
  \bibinfo{author}{\bibfnamefont{W.}~\bibnamefont{Losert}},
  \bibinfo{journal}{Proc Natl Acad Sci USA} \textbf{\bibinfo{volume}{106}},
  \bibinfo{pages}{8209} (\bibinfo{year}{2009}).

\bibitem[{\citenamefont{Peixoto}(2010)}]{Peixoto:2010}
\bibinfo{author}{\bibfnamefont{T.~P.} \bibnamefont{Peixoto}},
  \bibinfo{journal}{Phys Rev Lett} \textbf{\bibinfo{volume}{104}},
  \bibinfo{pages}{048701} (\bibinfo{year}{2010}).

\bibitem[{\citenamefont{Eldar and Eolwitz}(2010)}]{Eldar:2010}
\bibinfo{author}{\bibfnamefont{A.}~\bibnamefont{Eldar}} \bibnamefont{and}
  \bibinfo{author}{\bibfnamefont{M.~B.} \bibnamefont{Eolwitz}},
  \bibinfo{journal}{Nature} \textbf{\bibinfo{volume}{467}},
  \bibinfo{pages}{167} (\bibinfo{year}{2010}).

\bibitem[{\citenamefont{Lestas et~al.}(2010)\citenamefont{Lestas, Vinnicombe,
  and Paulsson}}]{Lestas:2010}
\bibinfo{author}{\bibfnamefont{I.}~\bibnamefont{Lestas}},
  \bibinfo{author}{\bibfnamefont{G.}~\bibnamefont{Vinnicombe}},
  \bibnamefont{and} \bibinfo{author}{\bibfnamefont{J.}~\bibnamefont{Paulsson}},
  \bibinfo{journal}{Nature} \textbf{\bibinfo{volume}{467}},
  \bibinfo{pages}{174} (\bibinfo{year}{2010}).

\bibitem[{\citenamefont{Glass and Kauffman}(1972)}]{Glass:1972}
\bibinfo{author}{\bibfnamefont{L.}~\bibnamefont{Glass}} \bibnamefont{and}
  \bibinfo{author}{\bibfnamefont{S.}~\bibnamefont{Kauffman}},
  \bibinfo{journal}{J Theor Biol} \textbf{\bibinfo{volume}{34}},
  \bibinfo{pages}{219} (\bibinfo{year}{1972}).

\bibitem[{\citenamefont{Braunewell and Bornholdt}(2009)}]{Braunewell:2009}
\bibinfo{author}{\bibfnamefont{S.}~\bibnamefont{Braunewell}} \bibnamefont{and}
  \bibinfo{author}{\bibfnamefont{S.}~\bibnamefont{Bornholdt}},
  \bibinfo{journal}{J Theor Biol} \textbf{\bibinfo{volume}{258}},
  \bibinfo{pages}{502 } (\bibinfo{year}{2009}).

\bibitem[{\citenamefont{Norrell et~al.}(2007)\citenamefont{Norrell, Samuelsson,
  and Socolar}}]{Norrell:2007}
\bibinfo{author}{\bibfnamefont{J.}~\bibnamefont{Norrell}},
  \bibinfo{author}{\bibfnamefont{B.}~\bibnamefont{Samuelsson}},
  \bibnamefont{and} \bibinfo{author}{\bibfnamefont{J.}~\bibnamefont{Socolar}},
  \bibinfo{journal}{Phys Rev E} \textbf{\bibinfo{volume}{76}},
  \bibinfo{pages}{46122} (\bibinfo{year}{2007}).

\bibitem[{\citenamefont{Norrell and Socolar}(2009)}]{Norrell:2009}
\bibinfo{author}{\bibfnamefont{J.}~\bibnamefont{Norrell}} \bibnamefont{and}
  \bibinfo{author}{\bibfnamefont{J.}~\bibnamefont{Socolar}},
  \bibinfo{journal}{Phys Rev E} \textbf{\bibinfo{volume}{79}},
  \bibinfo{pages}{61908} (\bibinfo{year}{2009}).

\bibitem[{\citenamefont{Gehrmann and Drossel}(2010)}]{Gehrmann:2010}
\bibinfo{author}{\bibfnamefont{E.}~\bibnamefont{Gehrmann}} \bibnamefont{and}
  \bibinfo{author}{\bibfnamefont{B.}~\bibnamefont{Drossel}},
  \bibinfo{journal}{Phys Rev E} \textbf{\bibinfo{volume}{82}},
  \bibinfo{pages}{046120} (\bibinfo{year}{2010}).

\bibitem[{\citenamefont{Klemm and Bornholdt}(2005{\natexlab{a}})}]{Klemm:2005a}
\bibinfo{author}{\bibfnamefont{K.}~\bibnamefont{Klemm}} \bibnamefont{and}
  \bibinfo{author}{\bibfnamefont{S.}~\bibnamefont{Bornholdt}},
  \bibinfo{journal}{Proc Natl Acad Sci USA} \textbf{\bibinfo{volume}{102}},
  \bibinfo{pages}{18414} (\bibinfo{year}{2005}{\natexlab{a}}).

\bibitem[{\citenamefont{Klemm and Bornholdt}(2005{\natexlab{b}})}]{Klemm:2005b}
\bibinfo{author}{\bibfnamefont{K.}~\bibnamefont{Klemm}} \bibnamefont{and}
  \bibinfo{author}{\bibfnamefont{S.}~\bibnamefont{Bornholdt}},
  \bibinfo{journal}{Phys Rev E} \textbf{\bibinfo{volume}{72}},
  \bibinfo{pages}{055101} (\bibinfo{year}{2005}{\natexlab{b}}).

\bibitem[{\citenamefont{Braunewell and Bornholdt}(2007)}]{Braunewell:2007}
\bibinfo{author}{\bibfnamefont{S.}~\bibnamefont{Braunewell}} \bibnamefont{and}
  \bibinfo{author}{\bibfnamefont{S.}~\bibnamefont{Bornholdt}},
  \bibinfo{journal}{J Theor Biol} \textbf{\bibinfo{volume}{245}},
  \bibinfo{pages}{638 } (\bibinfo{year}{2007}).

\bibitem[{\citenamefont{Shmulevich and Kauffman}(2004)}]{Shmulevich:2004}
\bibinfo{author}{\bibfnamefont{I.}~\bibnamefont{Shmulevich}} \bibnamefont{and}
  \bibinfo{author}{\bibfnamefont{S.}~\bibnamefont{Kauffman}},
  \bibinfo{journal}{Phys Rev Lett} \textbf{\bibinfo{volume}{93}},
  \bibinfo{pages}{48701} (\bibinfo{year}{2004}).

\bibitem[{\citenamefont{Kauffman}(1993)}]{Kauffman:1993}
\bibinfo{author}{\bibfnamefont{S.~A.} \bibnamefont{Kauffman}},
  \emph{\bibinfo{title}{The Origins of Order}} (\bibinfo{publisher}{Oxford
  University Press, New York}, \bibinfo{year}{1993}).

\bibitem[{\citenamefont{Langton}(1990)}]{Langton:1990}
\bibinfo{author}{\bibfnamefont{C.~G.} \bibnamefont{Langton}},
  \bibinfo{journal}{Physica D} \textbf{\bibinfo{volume}{42}},
  \bibinfo{pages}{12 } (\bibinfo{year}{1990}).

\end{thebibliography}
\end{document}